\documentclass[preprint,12pt]{elsarticle}
\journal{Computer Physics Communications}
\usepackage[T1]{fontenc} % if needed

\usepackage{url}
\usepackage{graphicx}
\usepackage{axodraw2}
\usepackage{color}
\usepackage{pstricks}
\newcommand{\be}{\begin{equation}}
\newcommand{\ee}{\end{equation}}
\newcommand{\bea}{\begin{eqnarray}}
\newcommand{\eea}{\end{eqnarray}}
\newcommand{\dd}{\mbox{d}}

\newcommand{\ep}{\varepsilon}

\begin{document}

\begin{frontmatter}

\title{
\normalfont
\vskip-4cm{\baselineskip14pt
  \begin{flushleft}
      \normalsize TTP13-047
  \end{flushleft}}
  \vskip4.5cm
  FIESTA 4: optimized Feynman integral calculations with GPU support}
%\title{FIESTA4: optimized Feynman integral calculations with GPU support}

\author[SRCC,KIT]{A.V.~Smirnov\corref{cor1}}
\ead{asmirnov80@gmail.com}

\cortext[cor1]{Corresponding author}

\address[SRCC]{Research Computing Center, Moscow State University, 119992\\ Moscow, Russia}
\address[KIT]{Institut f\"ur Theoretische Teilchenphysik, Karlsruhe Institute of Technology, \\ D--76128 Karlsruhe, Germany}

\begin{abstract}
This paper presents a new major release of the program  FIESTA
(Feynman Integral Evaluation by a Sector decomposiTion Approach).
The new release is mainly aimed at optimal performance at large scales 
when one is increasing the number of sampling points in order to 
reduce the uncertainty estimates.
The release now supports graphical processor units (GPU) for the numerical integration,
methods to optimize cluster-usage,
as well as other speed, memory, and stability improvements.
\end{abstract}
\end{frontmatter}
\newpage

{\bf PROGRAM SUMMARY}

\vspace{1cm}

\begin{small}
\noindent
{\em Manuscript Title:} FIESTA4: optimized Feynman integral calculations with GPU support\\
{\em Authors:} A.V. Smirnov\\
{\em Program title:} FIESTA4\\
{\em Licensing provisions:} GPLv2\\
{\em Programming language:} {\tt Wolfram Mathematica} 7.0 or higher, {\tt c++}\\
{\em Computer(s) for which the program has been designed:} from a desktop PC to a supercomputer\\
{\em Operating system(s) for which the program has been designed:} Unix, Linux, \\ Mac OS X\\
{\em RAM required to execute with typical data:} depends on the complexity of the \\ problem  \\
{\em Has the code been vectorized or parallelized?:} yes\\
{\em Number of processors used: } from 1 processor up to loading a supercomputer (tests were performed up to 2048 cores); from a personal GPU up to professional GPUs at a supercomputer \\
{\em Supplementary material:} The article, usage instructions in the program package, http://science.sander.su, https://bitbucket.org/fiestaIntegrator/fiesta/overview
{\em CPC Library Classification:} 4.4 Feynman diagrams, 4.12  Other Numerical
Methods, 5 Computer Algebra, 6.5 Software including Parallel Algorithms\\
{\em External routines/libraries used:} {\tt Wolfram Mathematica} [1], {\tt KyotoCabinet} [2], {\tt Cuba} [3], {\tt QHull} [4] \\
{\em Nature of problem:}
The sector decomposition approach to evaluating Feynman integrals
falls apart into the sector decomposition itself, where one
has to minimize the number of sectors; the pole resolution
and epsilon expansion; and the numerical integration of the resulting expression.
Morover, in cases where the integrand is complex, one has to perform a contour deformation.\\
{\em Solution method:}
The program has a number of sector decomposition strategies. Everything except the integration is performed in {\tt Wolfram Mathematica} [1] 
(required version is 7.0 or higher). This part of the calculation is parallelized with the use of shared memory.
The database is stored on hard disk with the use of the {\tt KyotoCabinet} [2] database engine.
\\
The integration part of the algorithm can be performed on a cluster. It is written in {\tt c++} and does not need 
{\tt Wolfram Mathematica}. For integration we use the Cuba library package [3].
\\
The sampling point evaluation has been vectorized. It can also use graphical processor units for the parallelization 
of sampling point evaluation.
\\
{\em Restrictions:} The complexity of the problem is mostly restricted
by CPU time required to perform the integration and obtain a proper precision.\\
{\em Running time:} depends on the complexity of the problem.\\
{\em References:} 
{\\} [1] http://www.wolfram.com/mathematica/, commercial algebraic software; 
{\\} [2] http://fallabs.com/kyotocabinet/, open source; 
{\\} [3] http://www.feynarts.de/cuba/, open source; 
{\\} [4] http://www.qhull.org, open source.

\end{small}

\newpage

\section{Introduction}

Sector decomposition is an automatic approach to the evaluation of Feynman integrals
initially introduced by Binoth and Heinrich~\cite{Binoth:2000ps,Binoth:2003ak,Binoth:2004jv,Heinrich:2008si,Bogner:2007cr,Bogner:2008ry}.

Feynman integrals have the following form:

\bea
  \mathcal F(a_1,\ldots,a_n) &=&
  \int \cdots \int \frac{\dd^d k_1\ldots \dd^d k_l}
  {E_1^{a_1}\ldots E_n^{a_n}}\,,
  \label{eqbn-intr}
\eea
where the denominator factors $E_i \equiv E_i - i0$ are linear functions with respect to
scalar products of loop momenta $k_i$ and external momenta $p_i$, $a_i$ are integers and dimensional regularization with
$d=4-2\epsilon$ is implied.

If one substitutes values for all kinematic invariants and masses, the integral can be evaluated numerically in the epsilon expansion.
The approach is based on the alpha-representation of Feynman integrals:

\begin{eqnarray}\label{Alpha}
    &&\mathcal F(a_1,\ldots,a_n) =(i\pi^{d/2})^l\times
   \\\nonumber
    &&\frac{\Gamma(A-l d/2)}{\prod_{j=1}^n \Gamma(a_j)}
            \int_{x_j\geq 0} d x_i\ldots d x_{n} \delta\left(1-\sum_{i=1}^n x_i \right)
                \left(\prod_{j=1}^n x_j^{a_j-1}\right) \frac{U^{A-(l+1)d/2}}{(F-i0)^{A-ld/2}},                         
\end{eqnarray}
where $A=\sum_{i=1}^n a_n$, $l$ is the number of loops and
$U$ and $F$ are constructively defined polynomials of $x_i$ (Symanzik polynomials, see, for example~\cite{Smirnov:2012gma}).

The sector decomposition approach has been implemented in three public programs --- {\tt sector\_decomposition} by Bogner and Weinzierl~\cite{Bogner:2007cr,Bogner:2008ry}, {\tt SecDec} by Binoth and Heinrich~\cite{Heinrich:2008si} 
(later improved and made public by Borowka, Carter and Heinrich~\cite{Carter:2010hi,Borowka:2012yc,Borowka:2012ii,Borowka:2012rt,Borowka:2013cma,Borowka:2013lda,Borowka:2014koa,Borowka:2015mxa}) and {\tt FIESTA}~\cite{Smirnov:2008py,Smirnov:2009pb,Smirnov:2013eza}. Here we present a new major release of {\tt FIESTA}.

The main goal of the current release is performance. After a number of algebraic transformations {\tt FIESTA} creates a database with multi-dimensional integrands, that have to be evaluated numerically.  One of the main disadvantages of the numerical approach is that in complex situations the error estimate of the result is rather high, and the only way to improve the answer is to increase the number of sampling points. Roughly speaking, a 100-time increase in the number of sampling points results in a 10-time decrease of the error estimate, hence one more reliable digit. However the time needed for such a gain also increases about 100 times (we ignore the algebraic preparation time which is a constant). 

Therefore when going for a high precision in complicated cases one needs to think of modern methods to gain the performance increase. This can be achieved in different ways such as the internal vectorization of the code, cluster usage and GPU usage. All those approaches are presented in the new release as well of ways to improve stability when working with large scales (for example, the possibility to continue a job that crashed on a cluster for reasons not related with {\tt FIESTA}).

The new version of {\tt FIESTA} is about 2--4 times faster than the old one, and that the usage of a GPU 
can increase performance 2--4 more compared with the pure CPU mode if there are enough sampling points.

In section 2 we will describe, how parallelization is used in {\tt FIESTA} and in section 3 we will
provide a number of measurements showing the speedup of the new version. 
Section 4 shows how to install {\tt FIESTA} and describes in detail its usage.
For the main algorithm description it is better to refer to articles
on previous versions of {\tt FIESTA}~\cite{Smirnov:2008py,Smirnov:2009pb,Smirnov:2013eza}.

\section{Parallelization in FIESTA}

When dealing with modern computers one faces the fact that the speed of individual computer cores is not growing much for many years. Hence for having a productivity gain one needs to deal with parallelization in some way. {\tt FIESTA} uses different approaches to parallelization.

\subsection{Code vectorization}

Modern processors are capable of using SIMD (single instruction multiple data) vector instructions which operate on multiple values contained in one large register at the same time. Basically it means that a processor is ready to operate with 4 double-sized floating point numbers (``doubles'') at the same time if the processors and the operating system support AVX (advanced vector extensions). For example, multiplying 4 pairs of doubles takes exactly the same time as multiplying 1 pair of doubles if those numbers are properly aligned in memory. 

While moving from {\tt FIESTA3} to {\tt FIESTA4} one of the important goals was to vectorize the function evaluation code. Remember, the code of {\tt FIESTA} takes the integration input as a string, parses it and transforms to an internal form --- a sequence of quadripples (operator, first operand, second operand, result). Afterwards the code runs a cycle through all the operations evaluating the function. Due to the nature of the problem, the integrands are almost rational functions (containing a small number of logarithms in addition to rational operations). Therefore it becomes reasonable to turn to evaluating multiple sampling points at the same time so that vectorization can be enabled. To do that we switched to the new version of the Cuba library\cite{Hahn:2014fua}.

It is worth noticing that this change does not require anything from the user, it is simply an integration performance gain, that will be demonstrated on tables below.

\subsection{GPU usage}

Another direction in gaining performance is the usage of graphical processor units (GPUs). The nature of this approach is similar --- GPUs are capable of performing bunches of similar operations at the same time. If one compares single operations, the GPUs are much slower compared to CPUs. However, they can win due to the number of threads that can work in parallell with no performance loss. GPUs are also simpler than CPUs being only able to perform straightforward calculation instructions, they are usually referred as a device (when compared to the CPU host).

The basic way to work with GPUs is the following. One prepares a function that can be evaluated on a GPU (the so-called kernel), then loads initial data into the GPU memory, then creates a grid of GPU threads that run the same evaluation kernel on different parts of the initial data. Finally, the result is copied back to RAM. This approach is used in the new version of {\tt FIESTA}. 

However this is not that straightforward as the vectorization approach. First of all, one needs to have a discrete GPU card on the computer in use to gain this advantage. {\tt FIESTA} uses the CUDA computational package which means that it works only with NVidia GPU cards. Second, one has to install proper GPU drivers that have the possibility to use the GPU for calculations and not only for producing the image on the monitor. Third, not any GPU can result in a performance gain. The most important parameter when choosing a GPU is the number of so-called streaming multiprocessors (SM) in it. A standard laptop GPU has only 2 SM, while a modern gaming desktop GPU has 16 SM. This difference is extremely important, and one cannot benefit with {\tt FIESTA} performance from a laptop GPU with 2 SM.

As for the code complications, one of problems that had to be solved while working on {\tt FIESTA4} was optimizing the number of sampling points that are sent to the GPU for parallel evaluation. This choice is especially important while taking into account the structure of GPUs. The points for evaluation are organized in so-called blocks of threads; in {\tt FIESTA} we use block size equal to 128 units. Now a number of blocks can be sent to the streaming multiprocessor at the same time. An SM, depending on the GPU model, can work with up to 16 blocks of this size at the same time, but it turned out that the evaluation functions are complex enough and use too many so-called registers in the GPU, so loading the GPU totally is impossible even after all optimizations. So the optimal number of blocks was set to 8. Therefore {\tt FIESTA4} works with the number of streaming multiprocessors times 1024 points at the same time on a GPU. 

Of course one could send more sampling points. Choosing a multiple of such a number would only increase performance, but here one faces another restriction --- the GPU memory. Each point evaluation requires a number of intermediate results to be stored during the evaluation of simple operations. And since the integrands can be huge, this number can be big enough. Now when evaluating the function at different sampling points at the same time, one needs to take the product of the number of bytes needed for evaluation of one point and the number of points. And this number of points has to fit into the GPU memory (which is exactly the number that is usually printed when one buys a GPU). Therefore it is impossible to be blindly increasing the number of sampling points used at the same time. Moreover, even with the default number {\tt FIESTA} has to check each time whether it is needed to decrease the number of sampling points for a particular huge integrand (especially if multiple integrator processes are working simultaneously with the use of same GPU sharing its memory).

\subsection{Parallel integration}
\label{PI}

{\tt FIESTA} package contains different binaries. 
When the integration is performed, one of those binaries, {\tt CIntegratePool}, reads the integrands from the integration database (prepared in {\tt Wolfram Mathematica}) 
and uses threads creating an integration queue, while the integration binaries work on particular expressions. Normally it is worth to use the number of threads equal to the number of kernels on the computer in use to achieve maximum performance.

The question arises on whether to use the GPU by all threads if it is available on your computer. There is an option in {\tt FIESTA} to set how many of integration threads should be using the GPU. Perhaps it might make sense to experiment on a particular computer on how it is optimal to choose this option. Tests show that at least 4 threads should be using a GPU.

\subsection{Integrator parallelization}

The new version of the Cuba library can work with multiple threads or ``accelerators'' (for example, GPUs) at the same time. This feature was introduced by T~Hahn partially for the needs of {\tt FIESTA} in 2014\footnote{Many thanks to T.~Hahn for being always ready to develop the Cuba library for the needs of its users.}. The idea was to have multiple threads in the integration binary receiving bunches of points where the function has to be evaluated. Some of those threads, the ``accelerators'' are receiving points only in large bunches of proper size required for the GPU optimization. 

Unfortunately by default we had to switch this option off. As explained in section~\ref{PI}, {\tt FIESTA} has another layer of parallelization and normally it is easier to have one CPU or GPU core working on a single integrand. Hence by default the Cuba parallelization is not used (turning it on might lead to a slowdown for the reason that transferring sampling points arrays to the evaluator threads with shared memory is a rather big overhead).
Still this option can be turned on in cases of extreme optimization when one has access to more processor kernels than there are integration terms in the database.

\subsection{Cluster usage}

One more approach to parallelization is supercomputer (cluster) usage. This is also not a new feature of {\tt FIESTA4} but it is worth to remind how it is organized. 
The {\tt CIntegratePool} binary has an {\tt MPI} (message passing interface) version, {\tt CIntegratePoolMPI}.
When an {\tt MPI} job is started on a cluster, it launches a number of equal tasks on different nodes. However they have an API for communication, and one of those is chosen to be ``master''. It opens the database with integrands and distributes the jobs to the ``slave'' nodes and each of those launches the integration binary to perform the calculations.

The current version of {\tt FIESTA} has a number of features aimed at stability when working on a cluster. One of the basic options that should probably be used at a cluster is {\tt -continue} that leads to all intermediate results being saved in the output database. Then in case a job crashes or reaches its timeout, the code can afterwards continue from the moment it stopped at. This option is useful for lengthy calculations when the integration time is dominating and the slowdown due to output database disk synchronization is negligible.

Next, {\tt FIESTA} watches whether the slave jobs are able to initialize integrators. If some of those are initialized improperly (which can happen due to cluster errors), it continues with other jobs. Moreover, if some of the slave jobs become frozen during the integration, {\tt FIESTA} is capable of resubmitting their tasks to other nodes.

\vspace{1cm}

To summarize, it is important to notice that in principle {\tt FIESTA} has two efficient layers of parallelization. One of those is related to different integrands, that can be distributed to multiple processes, on a single computer or on a cluster. The other is the possibility for the integrator to work with multiple points simultaneously, by vectorization or by using a GPU. The approaches can be used at the same time to gain maximal performance.

\section{Benchmarks}

\begin{figure}[htb]
\begin{center}
\includegraphics[width=0.5\textwidth]{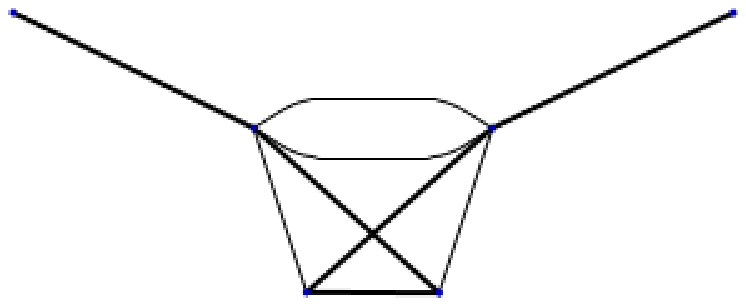}
\caption{Four-loop non-planar on-shell propagator diagram}
\end{center}
\label{int99}
\end{figure}

To demonstrate the speed gain in {\tt FIESTA4} we performed a number of tests. 
Our example is taken from the calculation of quark mass relations to four-loop order \cite{Marquard:2015qpa} (see fig.~1, thick lines are massive, thin lines are massless). 
The database size is about 200Mb and it contains 649 integrands.

The tests were performed on three different computers. The main difference in setups is the number of processor cores and the number of GPU cards.
(the first test has 4 cores and 1 GPU, second test --- 8 cores (at one processor) and 1 GPU and third --- 8 cores (two processors with 4 cores each) and 2 GPUs.

The first test was performed on a computer with an AMD Athlon(tm) II X4 640 Processor (4 cores, 3 GHz) and 
a ZOTAC GeForce GTX 580 AMP Edition (16 streaming multiprocessors, Fermi architecture).
We compared the GPU calculation, the CPU calculation and the old version of {\tt FIESTA} (CPU only) 
with different number of sampling points. The results are shown on table \ref{table99} (time is measured in seconds). 

\begin{table}[htb]
\begin{center}
\begin{tabular}{c|c|c|c}
Points   & GPU &  CPU & FIESTA3.4 \\\hline 
50 000    & $53$ & $51$ & $158$  \\\hline
500 000     &  $235$  &  $456$ & $1505$  \\\hline
5 000 000     &  $1819$ & $4243$ &  $16101$ 
\end{tabular}
\caption{Results and timings for the Feynman integral shown at fig.~1}
\label{table99}
\end{center}
\end{table}

We also made an optimization test for {\tt FIESTA} on a computer having an AMD Opteron(tm) Processor 6134 (8 cores, 2.3 Ghz)
and a Tesla X2070 GPU (14 streaming multiprocessors, Fermi architecture).
This kind of processors seems to be less optimal for the new features of {\tt FIESTA4}, so the speed gain is smaller.
It also turned out that the optimal approach is to have 4 cores use the GPU and
4 cores works with the CPU (the option {\tt gpu\_threads\_per\_node 4} was used, see section~4 for details).
The results are shown on table \ref{table99p}. Notation {\tt GPU 4/8} means that four out of eight cores per node use the graphical processors.

% Princeton. No access more

\begin{table}[htb]
\begin{center}
\begin{tabular}{c|c|c|c|c}
Points   & GPU4/8 & GPU  & CPU & FIESTA3.4\\\hline
50 000    & $39$ & $60$ & $46$ & $82$ \\\hline
500 000     &  $211$  & $255$  &  $423$ & $803$  \\\hline
5 000 000     &  $1772$ & $1992$ &  $4051$ & $8101$
\end{tabular}
\caption{GPU usage optimization for the Feynman integral shown at fig.~1}
\label{table99p}
\end{center}
\end{table}

A similar comparison was also performed on one of the nodes used in the Lomonosov supercomputer \cite{opanasenko2013lomonosov3827487}
with the MPI parallelization (however, the performance was measured on a single node).
Each node has 2 Intel Xeon X5570 processors (4 cores, 2.93 GHz) and 2 Tesla X2070 GPU cards (14 streaming multiprocessors, Fermi architecture).
We used the same database and tested the code with different number of sampling points.
The results can be seen in table \ref{table99L}. They demonstrate that having two GPUs can lead
to a speedup in 4 times compared to an 8-kernel node without GPUs
 (the option {\tt gpu\_per\_node 2} was used).
A certain slowdown for small numbers of sampling points is due to the overhead for the MPI communication.

\begin{table}[htb]
\begin{center}
\begin{tabular}{c|c|c|c}
Points   & GPU &  CPU & FIESTA3.4\\\hline
50 000    & $79$ & $70$ & $104$ \\\hline  
500 000     &  $133$  &  $410$ & $1063$ \\\hline  
5 000 000     &  $1065$ & $3770$ & $8451$   
\end{tabular}
\caption{Usage of 2 GPU for the Feynman integral shown at fig.~1}
\label{table99L}
\end{center}
\end{table}

%A similar calculation was perfomed on a larger integral contributing to the same evaluation.
%The database is about 1.1Gb and contains 9726 integrands. 
%e used the first computer listed in this section.

%\begin{table}[htb]
%\begin{tabular}{c|c|c|c}
%Points   & GPU &  CPU & FIESTA3.4 \\\hline
%50 000    & $1974$ & $2463$ & $6474$  \\\hline
%500 000     &  $11728$  &  $21479$ & $60435$  \\\hline
%\end{tabular}
%\caption{Results and timings for the diagram 9}
%\label{table9}
%\end{table}

The above examples show that the various parallelization models are is quite efficient. This makes it possible 
to achieve results which are out of reach with an approach without parallelization. While evaluating
master integrals involved in the calculation performed in \cite{Marquard:2015qpa} during a few months we used an amount of CPU times hours such
that a sequential evaluation would require about 1000 years on a single CPU.

\section{Installation and usage}

\subsection{Installation}

There are two ways to get {\tt FIESTA4} working: one can either download a binary package from 
\url{http://git.sander.su/fiesta/downloads} or build it from sources.
The binary package is not guaranteed to work at a particular machine. 
The binary packages with the current version of {\tt FIESTA} (4.0) were
build at Ubuntu 14.04LTS 64bit.
%and OS X 10.10 Yosemite 64bit.

Although {\tt FIESTA4} can be used without the {\tt c++} part (by setting {\tt UsingC} and {\tt UsingQLink} equal to {\tt False}),
it is not recommended since one will not have access to most features existing in this program (all the options
described in the previous section are related to the {\tt c++} part of {\tt FIESTA4})

To build {\tt FIESTA4} from sources one has to fullfill the following conditions.

1) The following libraries need to be installed on the computer:

\begin{itemize}
 
 \item {\tt MPFR} --- multiple-precision floating-point library that can be downloaded from \url{http://www.mpfr.org/} but has a big chance to exist in the system repositories. If one is downloading it from the official web-page and is not installing it system-wide, it is recommended to configure it with a prefix and install into a local directory. 
 {\tt FIESTA} is known to work with {\tt MPFR3.1.2};
 
 \item {\tt GMP} --- GNU multi precision library, it is also normally installed system-wide. If not, one can download it from \url{http://gmplib.org/}, configure and make. This library is not required at the object compilation step, and for linking one might point to the {\tt .libs} directory in the {\tt GMP} folder. {\tt FIESTA} is known to work with {\tt GMP5.1.3};
 
 \item In case one is going to use sector-decomposition strategies {\tt KU}, {\tt KU0}, {\tt KU2} or the {\tt SDExpandAsy} command, then one needs to have the {\tt qhull} convex hull search package installed. It might exist in the system repositories or can be downloaded from \url{http://www.qhull.org/}. {\tt FIESTA} is known to work with {\tt qhull2012.1}. A custom installation of {\tt qhull} will also require {\tt cmake} to be installed on your system;
 
 \item If one is going to build an {\tt MPI} version of {\tt CIntegratePool}, then one needs one of the {\tt MPI} environments to be installed on the system. Normally this is recommended if one is installing {\tt FIESTA} on a cluster;
 
 \item If one is building the GPU version of {\tt FIESTA}, the NVidia Cuda libraries and drivers (\url{https://developer.nvidia.com/cuda-downloads}) have to be installed.
 
\end{itemize}

2) Clone {\tt FIESTA} with {\tt git}

{\tt git clone https://bitbucket.org/feynmanIntegrals/fiesta.git}

3) Build from sources the libraries that {\tt FIESTA4} requires. 
{\tt FIESTA} comes with the database engine {\tt kyotocabinet 1.2.76} and Cuba library {\tt Cuba 4.0}.
Both codes were modified for some reasons so it is not recommended to download and install them manually
(for example, the Cuba library was modified to use the 64-bit integers for the number of sampling points
and to avoid the usage of shared memory by the integrator in case of one thread). 
Also the new release contains the {\tt uuid} library that is used by {\tt Mathlink} starting from {\tt Mathematica 10}.

{\tt make dep}

4) Build the {\tt FIESTA} binaries

{\tt make}

Do not forget to change the current directory to {\tt fiesta/FIESTA4} before giving those commands.
To build {\tt FIESTA} and the depending libraries faster, add {\tt -j n} to the {\tt make} commands
as usual, where {\tt n} is the number of kernels on the computer in use.

If the libraries mentioned earlier are installed and exist on the system paths, 
then one should simply {\tt cd} to the directory with {\tt FIESTA} and run the {\tt make} command. 
If not, one should first edit {\tt paths.inc} and add those paths with -I and -L for include and link paths correspondingly.

In rare cases one needs also to edit {\tt libs.inc} to tune library linking.
If one wants it static as much as possible, then {\tt libs.inc} should be replaced with {\tt libs-static.inc}
Normally one does not need to edit the Makefile, neither the main one, nor the ones in sub-folders.

The {\tt make} command should build everything but the {\tt MPI} version of CIntegratePool and the GPU version of the integration workers.
There is no {\tt make install} in the package.

If one is compiling {\tt FIESTA} at a cluster and is going to use only the {\tt C++ FIESTA}, it can be compiled with

{\tt make nomath}

To produce the {\tt MPI} integration pool binary one should run

{\tt make mpi}

The build results might be verified with {\tt make test} or {\tt make testall} (to test also {\tt MPI}). No errors should be produced.
These simple tests mainly test whether the binaries are functional.

To produce the GPU version of the integrators one should run

{\tt make gpu}

To get the latest development version of {\tt FIESTA}, one should switch to the {\tt dev} branch and compile {\tt FIESTA} again:

{\tt
git checkout dev
}

{\tt
make
}

In case a full rebuild is required (including the external libraries), one can call

{\tt
make cleandep \&\& make dep \&\& make clean \&\& make
}

{\tt FIESTA} is supposed to work at modern 64-bit distributions of {\tt Linux} and {\tt Mac OS X}.

\subsection{Code structure}

The main file that is loaded in {\tt Mathematica} is {\tt FIESTA4.m}. For not too lengthy calculations on personal computers it is enough to simply make {\tt Mathematica} calls that perform everything ``behind the scene''. For advanced usage it is better to understand the internal structure of {\tt FIESTA}.

The {\tt Mathematica} part prepares the integration databases (note: the databases differ depending on the {\tt UsingC} setting, so a database prepared for the {\tt c++} integration cannot be used by the pure {\tt Mathematica} integrator). Then if the {\tt c++} integrator is used, it launches the binary {\tt CIntegratePool} that evaluates the integrals in the database and uses temporary files to inform {\tt Mathematica} about results. {\tt Mathematica} reads those results and saves them in the output database. After everything is done, it uses the results in the output database to produce the final result.

An alternative approach is to run {\tt Mathematica} with the {\tt OnlyPrepare} option. Then it stops at the point when the integrand database is created and prints out the command that should be used in a shell to perform the integration. The advantage of this approach is that now one can use the integration command on different computers (without need to install {\tt Mathematica} there). Moreover, one can experiment with different integration options at this point without need to perform the algebraic part once again. Afterwards an output database is created and one can use the {\tt GenerateAnswer[]} command in {\tt Mathematica} to produce the final result.

The standard pool binary, {\tt CIntegratePool} uses threads to distribute tasks to integration workers. Another pool binary, {\tt CIntegratePoolMPI}, is an {\tt MPI} version that can be used at a cluster.

The pool binaries search for the integration workers in the same folder. There can be four versions of integration workers. All their names start with {\tt CIntegrateMP} and can have {\tt C} or {\tt G} letters added, where {\tt C} stands for complex. It is used if there is an integration with complex numbers. {\tt G} stands for the GPU version.

By default the CPU integration binary uses one CPU thread for integration, but this behavior can be changed. The GPU integration binary first searches for available GPUs. If they cannot be detected, it reports an error and stops. In other cases each binary by default will be using all available GPUs (launching the corresponding number of threads) and will not be using CPU integration. However the default behavior of the pool binary is to distribute different GPUs for the integration workers, so each of workers is using one GPU in the end.

\subsection{Program syntax}

This section mostly copies the one from the paper on {\tt FIESTA3} \cite{Smirnov:2013eza} but it should be here for consistency. Furthermore there are a few changes and new command variants.

{\tt FIESTA4} comes along with various examples that can be found in the directory {\tt examples}. 
All the examples listed in this section can be found in the file {\tt examples.nb}  in that directory.

If one is loading {\tt FIESTA} from {\tt Ma\-the\-ma\-tica}, it should either be loaded with
\newline
\vspace{0.1cm}{\tt SetDirectory[<path to FIESTA>]; Get["FIESTA4.m"];} \vspace{0.1cm}
\newline
or 
\newline
\vspace{0.1cm}{\tt FIESTAPath=<path to FIESTA>; Get[FIESTAPath<>"FIESTA4.m"];}.\vspace{0.1cm}
\newline
Do not load {\tt FIESTA} by simply specifying a full path to it, it will not work properly.

Now one can call the following commands:

\vspace{0.1cm}{\tt SDEvaluate[\{U,F,l\},indices,order]},\vspace{0.1cm}

\noindent where $U$ and $F$ are the functions from eq.~(\ref{Alpha}), $l$ is the number of loops,
$indices$ is the set of indices and $order$ is the required order of $\varepsilon$-expansion.

This command evaluates the Feynman integral with $(i\pi^{d/2})^l$ and $e^{-l \gamma_{\rm E} \epsilon}$ pulled out,
the same is true for the following commands.

To avoid manual construction of $U$ and $F$ one can use a build-in function {\tt UF}
and launch the evaluation as

\vspace{0.1cm}{\tt SDEvaluate[UF[loop\_momenta,propagators,subst],indices,order]},\vspace{0.1cm}

where $propagators$ are $(-E_i)$ from Eq. (\ref{eqbn-intr}), $subst$ is a set of substitutions for external momenta and masses 
(remember: the code performs numerical integration so the function $F$
should not depend on anything external).

\textit{Example}:  

\vspace{0.1cm}{\tt SDEvaluate[UF[\{k\},\{-k$^2$,-(k+p$_1$)$^2$,-(k+p$_1$+p$_2$)$^2$,-(k+p$_1$+p$_2$+p$_4$)$^2$\},
\\
\{p$_1^2\rightarrow$0,p$_2^2\rightarrow$0,p$_4^2\rightarrow$0,
p$_1$ p$_2\rightarrow$-S/2,p$_2$ p$_4\rightarrow$-T/2,p$_1$ p$_4\rightarrow$(S+T)/2,
\\
S$\rightarrow$3,T$\rightarrow1$\}],
\{1,1,1,1\},0]
}\vspace{0.1cm}

performs the evaluation of the massless on-shell box diagram (fig.~2).

\begin{figure}[htb]
\begin{center}
\includegraphics[width=0.5\textwidth]{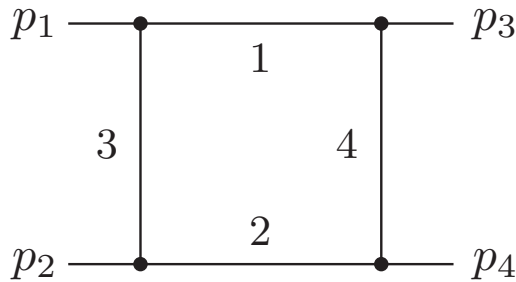}
\caption{Massless box diagram}
\end{center}
\label{box}
\end{figure}

The answer comes out in the following form:

\vspace{0.1cm}{\tt -4.38658 + (1.3333 + 0.000019 pm7)/ep\^{}2 \\ - (0.732466+0.000667 pm8)/ep + 0.001 pm9}\vspace{0.1cm}           

Here the terms with {\tt pm} are error estimates. {\tt FIESTA} collects the error estimates produced by the integrator 
and sums them using the mean-square norm. We recommend to treat this number as the $\delta$ value (standard deviation) in the normal representation
(hence it might make sense to multiply it by $3$).
If an error estimate is missing for some term, it means that is is smaller than $10^{-6}$.
{\tt FIESTA4} is more accurate than the previous versions in generating small error estimates
(see options {\tt DigitsLimit} and {\tt SmallNumberMultiplyers} described below for details); {\tt FIESTA3} 
would not produce the {\tt pm7} and {\tt pm8} in the above example.

There can be one more argument to {\tt SDEvaluate} designating a variable. In case it exists, 
function {\tt F} is differentiated by this variable, and then $0$ is substituted.

In the following commands we will only provide the first version of the syntax (with {\tt \{U,F,l\}}). However, in all places this triple can be replaced with the {\tt UF} generator.

If one needs to  expand a Feynman integral by a small parameter {\tt tt} one should use

\vspace{0.1cm}{\tt SDExpand[\{U,F,l\},indices,order,tt,order in tt]}.\vspace{0.1cm}

\textit{Example}:  

\vspace{0.1cm}{\tt SDExpand[\{x[1] + x[2] + x[3] + x[4], x[1] x[3] + t x[2] x[4], 1\}, \{1, 
  1, 1, 1\}, 1, t, 1] }\vspace{0.1cm}

In this box example expansion one gets the following answer:

\vspace{0.1cm}
{\tt
0.000025 + 0.000343 pm346 + ep (-11.8703 + 0.010594 pm347)\\ + 4./(
 ep$^2$ t) + (-13.1596 + 0.000102 pm337)/t + (
 ep (-13.6234 + 0.001227 pm338))/t + (-0.00001 + 0.000135 pm352) t + 
 ep (5.18514 + 0.005176 pm353) t + 
 ep (1.99993 + 0.001143 pm349) Log[t] - (2. Log[t])/(ep t) + (
 ep (11.5147 + 0.000091 pm341) Log[t])/t \\ + 
 ep (-0.499977 + 0.001654 pm355) t Log[t] + 
 ep (-0.999987 \\ + 0.000015 pm350) Log[t]$^2$ - (4.*10$^{-6}$ ep Log[t]$^2$)/t \\ +
  ep (0.500017 + 0.000073 pm356) t Log[t]$^2$ \\ + (0.333332 ep Log[t]$^3$)/t
}
\vspace{0.1cm}

There is another way to expand this integral with the regions approach, however one has to provide
a regularization variable with the {\tt RegVar} setting and shift indices (for details see \cite{Smirnov:2013eza}). 
One cannot predict which indices have to be shifted; if one does not shift enough indices,
{\tt FIESTA} might produce error messages and fail.

\textit{Example}:  

\tt
\vspace{0.1cm}RegVar=la;

SDExpandAsy[\{x[1] + x[2] + x[3] + x[4], x[1] x[3] \\ + t x[2] x[4], 1\}, \{1, 
  1, 1, 1\}, 1, t, 1]\vspace{0.1cm}

\normalfont

In this case the integral is evaluated analytically and the answer is

{\tt
\vspace{0.1cm}
4/(ep$^2$ t) - (4 $\pi ^2$)/(3 t) - (2 Log[t])/(ep t) + 
 ep (-2 - $\pi^2$  \\ + 2 Log[t] - Log[t]$^2$ + 
    t (1/4 + $\Pi^2$/2 - Log[t]/2 + Log[t]$^2$/2) + \\ (
    7/6 $\pi^2$ Log[t] + Log[t]$^3$/3 + 17/3 PolyGamma[2, 1])/t)
    \vspace{0.1cm}
}

Those answers looks quite different, but if one subtracts one from another and expands the result properly, it can be clearly seens that it is the same answer.

There is one command that makes possible to apply sectors decomposition to integrals different from Feynman integrals:

\vspace{0.1cm}{\tt SDEvaluateDirect[var,function,degrees,order,deltas\_optional]}.\vspace{0.1cm}

Here {\tt var} stands for the integration variable used in functions (for example, {\tt x} goes for {\tt x[1]}, {\tt x[2]}, ...),
{\tt functions} is a list of functions and {\tt degrees} is the list of their exponents. {\tt order} is the requested order is epsilon,
and {\tt deltas} goes for the list of delta functions attached to the integrand. By default is is empty. For example,
{\tt \{\{1,3\},\{2,4\}\}} goes for the product of {\tt Delta[x[1]+x[3]-1]} and {\tt Delta[x[2]+x[4]-1]}.

\textit{Example}:  

\vspace{0.1cm}\tt
SDEvaluateDirect[x, \{1, 
  x[1] x[2] x[3] + x[1] x[2] x[4] + 
  
  x[1] x[3] x[4] + x[1] x[2] x[5], 
  x[1] x[3] + x[2] x[3] + 
  
  x[1] x[4] + x[2] x[4] + x[3] x[4] + 
   x[1] x[5] + x[2] x[5] + 
   
   x[3] x[5]\}, \{1, -1 - 2 ep, -1 + 
   3 ep\}, 0, \{\{1, 2, 3, 4, 5\}\}]

\normalfont\vspace{0.1cm}

A similar syntax works for the expansion. In this example the integrator needs access to 
{\tt Mathematica} to evaluate a {\tt PolyGamma} function, so the path is passed
with the {\tt MathematicaBinary} argument:

\vspace{0.1cm}{\tt SDExpandDirect[var,function,degrees,expand\_var,\\ deltas\_optional]}.\vspace{0.1cm}

\textit{Example}:  

\vspace{0.1cm}\tt
MathematicaBinary="math";

SDEvaluateDirect[x, \{1, 
  x[1] x[2] x[3] + x[1] x[2] x[4] + 
  
  x[1] x[3] x[4] + x[1] x[2] x[5], 
  x[1] x[3] + x[2] x[3] + 
  
  x[1] x[4] + x[2] x[4] + t (x[3] x[4] + 
   x[1] x[5] + 
   
   x[2] x[5] +  x[3] x[5])\}, \{1, -1 - 2 ep, -1 + 
   3 ep\}, 0, t, 0, 
   
   \{\{1, 2, 3, 4, 5\}\}]\vspace{0.1cm}

\normalfont

The answer has the same format with the previous commands.

There is one more way to use {\tt FIESTA}. An analytic Feynman integral evaluation method suggested by Lee \cite{Lee:2009dh,Lee:2010cga,Lee:2011jt,Lee:2012te} needs to know
for which values of space-time dimension {\tt d} the integral can have poles. {\tt FIESTA} can locate those values with the use of the following command:

\vspace{0.1cm}{\tt SDAnalyze[\{U,F,l\},indices,dmin,dmax]}.\vspace{0.1cm}

Here {\tt dmin} and {\tt dmax} are the ends of the interval where poles should be located. This syntax used only algebraic transformations,
so the result is exact. However, the program might miss some pole cancellations, so some of the returned values might be ``false alerts''.

\textit{Example}:  

\vspace{0.1cm}{\tt SDAnalyze[UF[\{k\},\{-k$^2$,-(k+p$_1$)$^2$,-(k+p$_1$+p$_2$)$^2$,-(k+p$_1$+p$_2$+p$_4$)$^2$\},
\\
\{p$_1^2\rightarrow$0,p$_2^2\rightarrow$0,p$_4^2\rightarrow$0,
p$_1$ p$_2\rightarrow$-S/2,p$_2$ p$_4\rightarrow$-T/2,p$_1$ p$_4\rightarrow$(S+T)/2,
\\
S$\rightarrow$3,T$\rightarrow1$\}],
\{1,1,1,1\},1,8]
}\vspace{0.1cm}

Returned answer is {\tt \{2,4\}} which means that the integrand has poles for {\tt d} equal to $2$ and $4$.

\subsection{Options of Mathematica FIESTA}

\label{FIESTAoptions}

Most of the options are available when calling {\tt FIESTA} from {\tt Mathematica}. However for fine-tuning of performance and especially when integrating on a cluster or with GPU usage one should consider creating an integrand database first and then running the integration from a shell, directly accessing the pool binaries.

THe options used in {\tt Mathematica} are provided by giving values to the following variables before calling {\tt FIESTA} commands such as {\tt SDEvaluate}. None of the options are obligatory but one should pay attention to the {\tt Datapath} in case {\tt FIESTA} files are located on a network drive, to the parallelization options for performance and to the complex mode options in case a physical region is under consideration. The following options are listed alphabetically.

\begin{itemize}

 \item {\tt AnalyticIntegration}: an option used only for {\tt SDExpandAsy}, {\tt True} by default, tells {\tt FIESTA} to try to take some integrations analytically after introducing regions;

\item {\tt BisectionPoint}: see {\tt ComplexMode};

\item {\tt BisectionPoints}: see {\tt ComplexMode};

\item {\tt BisectionVariables}: see {\tt ComplexMode};

 \item {\tt BucketSize} ($25$ by default): an option tuning the database usage (for details see the documentation on KyotoCabinet); increasing this variable might result is faster database access, but increases the RAM usage; since this number is treated as the exponent, increasing it much can result in integer overflow and unexpected results;

\item {\tt CIntegratePath}: by default the integration pool chooses itself the integration binary, however one might provide another path. In this case the integral logic of using or not using the version with complex numbers or GPU/CPU usage is switched off, so one has to be careful in choosing a proper binary;

 \item {\tt ComplexMode} ({\tt False} by default): with this setting set to {\tt True} {\tt FIESTA} performs a contour deformation in order to avoid poles in physical regions. The deformation size depends on a parameter, that is either set manually by giving a value to the {\tt ComplexShift} variable or is tuned automatically in the interval from zero to {\tt MaxComplexShift} ($1$ by default). Increasing the option {\tt LambdaSplit} ($4$ by default) might result in better tuning but slows the preparation; same is true for the search option {\tt LambdaIterations} that is set by default to $1000$; if the {\tt ConservativeComplexShift} option is set to {\tt True}, we divide the finally obtained $\lambda$ by two (to be on the safe side);
 
 The contour transformation cannot deal with cases where $F$ turns to zero for the reason that some variables are equal to $1$ (for example, a factor of $x-1$). In order to trace those cases, set {\tt TestF1=True}. In order to handle those singularities, set the {\tt BisectionVariables} equal to the list of variables such that the integration cube is divided into two parts. By default, the separation point is equal to $1/2$, however this can be changed either by setting the {\tt BisectionPoint} value, or by providing a list of {\tt BisectionPoints};

\item {\tt ComplexShift}: see {\tt ComplexMode};
 
 \item {\tt ConservativeComplexShift}: see {\tt ComplexMode};

 \item {\tt CPUCores}: an alternative way to parallelize the integration by making each integration worker use multiple threads. By default each integration worker uses one core, but it can be changed with this option. Normally {\tt NumberOfLinks} is more efficient, but there might be situations when increasing {\tt CubaCores} leads to better results;
 
  \item {\tt CurrentIntegrator} ({\tt vegasCuba} by default): the integrator used at the final stage. The options allowed in the current setup are {\tt vegasCuba}, {\tt suaveCuba}, {\tt divonneCuba} and {\tt cuhreCuba}. In principle, it is also possible to add ones own integrators by modifying the {\tt integrators.c} source file, however this is far beyond the standard usage of {\tt FIESTA}. The related parameter ({\tt CurrentIntegratorOptions}) presents a list of options of the currently chosen integrator (for details see \cite{Hahn:2004fe}). By default it has no value and the actual options are printed out when one starts the evaluation (the defaults are stored within the {\tt c++} part). The most commonly used integrator option is {\tt maxeval}, which can be set, for example, by {\tt CurrentIntegratorOptions = \{\{"maxeval","500000"\}\}} which sets the maximal number of sampling point for an integral. The default value is {\tt 50000}. For details of integrator options please refer to \cite{Hahn:2004fe}.

 One more virtual integrator (possible value of {\tt CurrentIntegrator})  is {\tt justEvaluate}. It simply evaluates the integrand at a given point, by default it is the point where all coordinates are equal to $1/2$. Its options are ${\tt x1, x2}$ and so on representing the integration coordinates;
 
  \item {\tt d0} ($4$ by default): specifies the integer space-time dimension;

 \item {\tt DataPath}: by default {\tt FIESTA} stores databases in the {\tt temp} sub-folder (relative to the current folder), their names start with {\tt db}. However one might wish to direct it elsewhere, especially if the folder with {\tt FIESTA} is on a network disk. The database should be preferably stored on a fast local disk;
 
  \item {\tt DigitsLimit}: the maximal number of digits produced in results, default is $6$; this option only influences result output but not calculations; 
 
  \item {\tt ExactIntegrationOrder}: (if set) specifies the order in epsilon where {\tt FIESTA} tries exact integration for some time (with the {\tt Mathematica} {\tt Integrate} command). The default time is $10$ seconds per sector and can be modified by the {\tt ExactIntegrationTimeout} option; works only with {\tt UsingC} set to {\tt False};
 
  \item {\tt FixSectors}: ({\tt True} by default) --- for the reason of fixing sector numbers and easier debugging we perform the variable substitution stage on the main kernel. If one sets this option to false, this stage will be also made parallel, however it normally does not influence the total time much;
 
  \item {\tt GPUIntegration}: makes the integration pool binary use GPU integration workers;
 
 \item {\tt LambdaIterations}: see {\tt ComplexMode};  
 
 \item {\tt LambdaSplit}: see {\tt ComplexMode};

 \item {\tt MathematicaBinary}: a path to the executable {\tt Mathematica} kernel. If set, it is passed to the integration pool, so it can request {\tt Mathematica} for values of functions it cannot evaluate itself (currently this feature is used only for {\tt PolyGamma});
 
 \item {\tt MaxComplexShift}: see {\tt ComplexMode};
 
  \item {\tt MixSectors} ($0$ by default): lets {\tt FIESTA} to sum up integrands in different sectors before integration; 
 
  \item {\tt NegativeTermsHandling} --- the mode used during the pre-resolution; default is {\tt Auto} which is equivalent with using {\tt AdvancedSqu\-ares3} while {\tt ComplexMode} is {\tt False} and {\tt None} otherwise. Possible values are {\tt None}, {\tt Auto}, {\tt Squares}, {\tt AdvancedSquares}, {\tt AdvancedSquares2} and {\tt Advanced\-Squares3}; for details see \cite{Smirnov:2009pb};
 
  \item {\tt NoDatabaseLock}: prevents {\tt FIESTA} from locking the database. This may be a way to avoid restrictions on some file systems but might result in corrupted databases;
 
 \item {\tt NumberOfSubkernels}: the number of subkernels that {\tt Mathematica} launches. Might be set equal to the number of cores on the computer in use but should not exceed the number of licensed subkernels. This setting can speed up the integrand preparation;
 
 \item {\tt NumberOfLinks}: the number of {\tt CIntegrate} processes that will be launched. The name of this option is left for compatibility with old versions of {\tt FIESTA} where each {\tt CIntegrate} process was called by a separate {\tt MathLink} connection. This setting corresponds to the parallelization during integration;
 
 \item {\tt OnlyPrepare}: by default this option is set to {\tt False}. In this case the calculation is performed completely, otherwise a database is prepared for integration and a shell command to run {\tt CIntegrate} without {\tt Mathematica} is printed. This can be used if one is preparing a database on one computer and is integrating elsewhere, or if one is willing to try different integrators or precision requests;
 
 \item {\tt PolesMultiplicity}: an option used only for {\tt SDAnalyze}, {\tt False} by default, changes the answer so that it returns not only values of {\tt d} but also maximal pole multiplicities;
 
  \item {\tt PreResolve}: an option used only for {\tt SDExpandAsy}, {\tt False} by default, tells {\tt FIESTA} to use the pre-resolution before searching for regions;
 
  \item {\tt PrimarySectorCoefficients}: one might specify the list of coefficients before primary sectors. A zero means that this sector will be ignored. With this setting one can split the problem into parts and also take diagram symmetries into account;
 
 \item {\tt QHullPath}: if one uses strategies {\tt KU}, {\tt KU0}, {\tt KU2} or the new {\tt SDExpandAsy} syntax, a correct path to the {\tt qhull} executable should be provided. By default it is set to {\tt qhull} assuming that the package is installed on the system, however one might provide a specific path;

  \item {\tt RegVar}: an option introduced for {\tt SDExpandAsy} but usable also in other situations. Sets an extra regularization variable;
 
  \item {\tt RemoveDatabases} ({\tt True} by default): specifies whether the database files should be removed after the integration;

 \item {\tt ReturnErrorWithBrackets}: ({\tt False} by default) changes the output --- with {\tt True} the error estimates are printed as {\tt pm[NUMBER]} instead of {\tt pmNUMBER};
 
 \item {\tt SeparateTerms}: if {\tt True}, the integrator receives integrable terms independently, not whole expressions for each sector; on the one hand, this simplifies the integrands and the integrators return better precision. On the other hand the error grows after summing up the results, so normally there is no recommendation on whether to use this option or not. However, in the {\tt MPI} mode this option might lead to a significant speedup since in this case it leads to a better parallelization;
 
  \item {\tt SmallNumberMultiplyers}: two numbers used to remove what is believed to be zero from the result; {\tt FIESTA} cuts out a number either \\
 i) its absolute value is smaller that {\tt 10} in the power {\tt -DigitsLimit} times the first number or\\
 ii) its absolute value is smaller that {\tt 10} in the power {\tt -DigitsLimit} times the second number and the absolute value of the result is smaller than {\tt 3} times the absolute value of the error estimate. The default numbers are {\tt \{3, 300\}}.
 
 \item {\tt SmallX}, {\tt MPThreshold}, {\tt PrecisionShift}, {\tt MPPrecision}, {\tt MPMinb}: the options that allow to fine-tune the MPFR subsystem. For details see the publication on {\tt FIESTA2} \cite{Smirnov:2009pb};
 
 \item {\tt STRATEGY}: sector decomposition strategy. By default we use {\tt STRATEGY\_S} \cite{Smirnov:2008py}, but there are also such options as {\tt STRATEGY\_B} \cite{Bogner:2007cr}, {\tt STRATEGY\_X} \cite{Binoth:2003ak}, {\tt STRATEGY\_SS} \cite{Speer:1968:AR} and {\tt STRATEGY\_KU}, {\tt STRATEGY\_KU0}, {\tt STRATEGY\_KU2}. Among the last three strategies the last variant is the full implementation of the algorithm from \cite{Kaneko:2009qx}, the first two are faster but might result in more sectors;

 \item {\tt TestF1}: see {\tt ComplexMode};
 
  \item {\tt UsingC}: by default this option is set to {\tt True}. This means that {\tt FIESTA} uses the {\tt c++} integration. If set to {\tt False}, it switches to {\tt Mathematica} integration, however this is not recommended;

 \item {\tt UsingQLink}: by default this option is set to {\tt True}. Switching it off will turn off database usage, however in {\tt FIESTA4} this is possible only together with {\tt UsingC=False};

 \item {\tt FastASY}: ({\tt False} by default, works only if {\tt PreResolve} is set to {\tt False}) specifies the region search mode (used in {\tt SDExpandAsy}). With {\tt FastASY} set to {\tt False} the polynomial {\tt U $\times$ F} is analyzed, with {\tt True} only the $F$ polynomial. The {\tt FastASY} variant might work significantly faster and will produce correct results almost all the time, but one should use it at his/her own risk;

\end{itemize}

The following options from {\tt FIESTA3} are no longer supported: {\tt CubaBatch}, {\tt CubaCores}.

\subsection{Options of the pool binaries}

\label{PoolOptions}

The {\tt CIntegratePool} and {\tt CIntegratePoolMPI} binaries can be used directly to perform the integration.
A big part of those options can be used from {\tt Mathematica}, but
if one already has a database, the binary has to be called directly. 
Moreover this gives a possibility to optimize performance by
trying different options without the need to perform the algebraic part again and to be safe from system
crashes by being able to continue from almost the place when it crashed (by setting appropriate options).

\begin{itemize}
 \item {\tt -in}: provides the path to the database with integrands (the .kch suffix can both be omitted or be there with the new version);
 \item {\tt -out}: provides the path to the database where results are stored (if the input database path ends with in or in.kch, this path can be omitted and will be automatically generated by replacing in.kch with out.kch);
 \item {\tt -from\_mathematica}: instructs {\tt CIntegratePool} that it was called from {\tt Mathematica}, so that it does not save the results in the output database, but uses temporary files in order to transfer results back and does not print the results to {\tt stdout}; should normally not be called from the console;
 \item {\tt -math}: provides the path to the {\tt Mathematica} binary; required only if the integrator needs access to polygamma evaluation;
 \item {\tt -bucket}: provides the bucket value for the output database; default is 10 (without additional options the output database has only a few entries); should be possibly increased if all integral results are saved;
 \item {\tt -CIntegratePath}: provides the path to the {\tt CIntegrate} binary. This can be either a full path (starting with {\tt /}) or just a filename, in this case {\tt CIntegratePool} searches for this file in the same directory. If this option is missing, it searches for {\tt CIntegrateMP}, {\tt CIntegrateMPC}, {\tt CIntegrateMPG} or {\tt CIntegrateMPCG} depending on whether the complex mode is on or off and whether the GPU usage is on or off;
 \item {\tt -integrator}: sets the integrator to be used. {\tt -intpar} sets some integrator parameter, for example, {\tt -intpar maxeval 500000}. For the list of integrators see section~\ref{FIESTAoptions};
 \item {\tt -MPThreshold}, {\tt -MPPrecision}, {\tt -PrecisionShift}, {\tt -SmallX}, {\tt -MPMin}: options that are fine-tuning the {\tt MPFR} subsystem;
 \item {\tt -threads}: sets the number of {\tt CIntegrate} processes launched by the pool. This option is meaningless in the {\tt MPI} mode;
 \item {\tt -testF}: instead of the integration, the code checks whether the sign of the imaginary part of $F$ is negative. Might be used for debugging special cases in complex mode, for details see \cite{Smirnov:2013eza};
 \item {\tt -debug}: used to print all integration results.
 \item {\tt -mpfr}: evaluate all integration points with the use of MPFR; takes much longer, but can be used from debugging;
 \item {\tt -complex}: specifies that the expression is complex. If one knows it do be real, this setting should not be used since it can slow down the integration a lot;
 \item {\tt -test}: perform an integrator test only (verifies whether the integrator works correctly on a sample linear function), {\tt -test\_integrator}: perform this test before the actual start, {\tt -preparse}: perform parse check of all expressions before integration;
 \item {\tt -task} followed by a number instruct the code to evaluate only expression related to one {\tt SDEvaluate} call. Normally, there is only one task in the database, so one would call {\tt -task 1}, but the {\tt SDExpandAsy} mode uses multiple tasks.
 \item {\tt -prefix} can be used to tell the program to integrate only with given powers of $\ep$ and {\tt RegVar}. 
    For example {\tt -task 1 -prefix  "\{-2,-\{1, 2\}\}"} corresponds to integrals having $\ep$ order $-2$ and {\tt RegVar} coefficient {\tt RegVar * Log [RegVar]\^{ }2};
    in case of no dependance on the {\tt RegVar} a simpler syntax like {\tt -prefix 3} can be used meaning that the integrals having $\ep$ order $3$ have to be calculated;
 \item {\tt -only\_prepare} only creates the output database with dummy entries so that it can be used with the {\tt GenerateAnswer[]} command; the entries contain print warnings that will appear on the screen if called from {\tt Mathematica}; this can be useful if one is running tasks with different prefixes one after another (for example, to bypass time restrictions on a cluster)
 \item {\tt -part} followed by two numbers, for example, {\tt 2/5}; makes the code only use the second part of five parts of the integrands of the current $\ep$ order; can be used in consequent runs for different parts, and the code will be summing up those results in the output database;
 \item {\tt -continue}: another option for bypassing time restrictions; makes the code save results for individual integrals in the output database; as soon as all the integrals for a given order are ready, the code calculates the results for this order and deletes the intermediate entries from the database; this mode and {\tt -save\_all} might require a larger {\tt bucket} value (minimum {\tt Log(2,n)-1}, where {\tt n} is the number of integrands);
 \item {\tt -save\_all}: same with {\tt -continue}, but the intermediate results are not erased;
 \item {\tt -separate\_terms}: instructs the algorithm not to group expressions by sectors and to integrate each integrable term separately, might be useful if one is using massive {\tt MPI} parallelization;
 \item {\tt -gpu}: makes the pool binary use GPU integration workers;
 \item {\tt -gpu\_threads\_per\_node}: limits the number of workers that will use GPU; default value is 0 which means no restriction, however the pool tries to distribute one GPU per worker; the value -1 means that each worker will use all available GPUs;
 \item {\tt -gpu\_per\_node}: instructs the pool binary how many GPUs are to be used per node (or at this computer in case of threads); this option is important in case of multiple GPU so that they are properly distributed by the integration workers and so that the workers know that they are not the sole users of GPUs and use only an appropriate part of the GPU memory.
 \item {\tt -cpu\_threads\_per\_worker}: sets the number of CPU threads to be used by each integration worker; default is 1 which means 1 thread (and none in case of GPU usage); this is an alternative parallelization option;
 \item {\tt -print\_command}: prints the command used to call the pool binary on stdout; useful when dealing with multiple databases and logs;
\end{itemize}

The following options were removed when compared with {\tt FIESTA3}: {\tt all}, {\tt direct}, {\tt nopreparse}, {\tt notest}, {\tt CubaBatch}, {\tt CubaCores}.

\subsection{CIntegrate options}
\label{CIntegrate}
Individual integrals are evaluated by the integration worker binaries. As it was said earlier, {\tt FIESTA} comes with four of those binaries:
{\tt CIntegrateMP}, {\tt CIntegrateMPC}, {\tt CIntegrateMPG} and {\tt CIntegrateMPG}, where {\tt MP} stands for internal multi-precision evaluations (the current version of {\tt FIESTA} no longer has binaries without MPFR), {\tt C} stands for complex and {\tt G} stands for the GPU usage.

It is also possible to use an external integrator if it supports the protocol explained below. Moreover, those integrator workers can be used separately without {\tt FIESTA}.
Each binary has no options on start, accepts input from {\tt stdin} and print it to {\tt stdout} (alternatively if the call has one argument it is treated as input file name). Each command sent to the program is ended with a new line symbol.

The main command to be provided to the program input is {\tt Integrate}. After that one should send the expression. It consists of a number of lines, each of them should be ended with the {\tt ;} symbol. At the end should be a line consisting of the {\tt |} symbol. The expression lines are the following:

\begin{itemize}
 \item The number of variables;
 \item The number of intermediate functions;
 \item A number of lines each representing an intermediate expression. If the second line is {\tt 0;}, then this part should be missing;
 \item The final expression.
\end{itemize}

The expressions might contain algebraic operations such as {\tt +, -, *, /}, bracket symbols, numbers with a floating point. The integration variables should be referred as {\tt x[1], x[2]} and so on, intermediate functions as {\tt f[1], f[2]} and so on. Power is represented as {\tt p[expr,exponent]}, natural logarithm as {\tt l[expr]}. One can also use {\tt P} for $\pi$ and {\tt G} for {\tt EulerGamma}. {\tt PolyGamma[arg1,arg2]} also works but one needs to provide a path to the {\tt Mathematica} binary --- the integrator cannot evaluate this function on its own. However a few values for small polygamma arguments are hard-coded 
({\tt PolyGamma[1,1], PolyGamma[2,1], PolyGamma[2,2], PolyGamma[2,3], PolyGamma[2,4], PolyGamma[3,1], PolyGamma[3,2]}).

\textit{Example}:  

\tt

1;

0;

x[1]+0.2;

|

\normalfont

If one feeds this example into the integrator program after the {\tt Integrate} command then it will result in integrating $x+0.2$ from $0$ to $1$.

The program also accepts the following commands (most of them require an argument passed as next line):

\begin{itemize}
 \item {\tt Parse}: same as {\tt Integrate}, but the expression is only parsed;
 \item {\tt Exit} or {\tt Quit}: quit the program; 
 \item {\tt SetMath}: provides a path to the {\tt Mathematica} binary;
 \item {\tt SetIntegrator}: sets the integrator to be used;
 \item {\tt SetCurrentIntegratorParameter}: sets one of the integrator parameters, the next line should be the parameter name, the line after that --- the value;
 \item {\tt GetCurrentIntegratorParameters}: simply returns the list of current parameters and their values;
 \item {\tt MPFR}, {\tt Native} and {\tt Mixed} (default variant): chooses whether the integrand should use {\tt MPFR} everywhere, the double precision or the mixed mode. The mixed mode used the following five options to determine in which parts of the integration cube which arithmetics should be used: {\tt SetMPPrecision}, {\tt SetMPPrecisionShift}, {\tt SetMPMin}, {\tt SetMPThreshold}, {\tt SetSmallX};
 \item {\tt Debug}: makes the code print values in all integration points;
 \item {\tt TestF}: instead of the integration the code checks the sign of the imaginary part of the integrand;
 \item {\tt Statistics} prints out different timing statistics of the program; to reset all the timers use {\tt ClearStatistics};
 \item {\tt CPUCores} sets the number of CPU threads that are used by the worker; default is $1$ or $0$ in case of GPU;
 \item {\tt GPUCores} sets the number of GPU threads that are used by the worker; default is $0$ for the CPU worker and the automatically detected number of graphical accelerators in case of the GPU worker;
 \item {\tt GPUForceCore} works only if {\tt GPUCores} is set to $1$ and forces the single evaluation core to use the GPU with the given number; the numbering is from $0$ to the number of accelerators minus one; providing any integer number is OK since the code takes the remainder from divining this number by the number of accelerators found on the computer;
 \item {\tt GPUMemoryPart} instructs the worker that it is not the single user of the current GPU and should be use only a part of the available memory provided by this number, default is $1$;
 \item {\tt GPUData} prints out the information about the GPUs that exist on the current computer; 
 \item {\tt Help}: lists all those commands.
\end{itemize}

Note: if the worker is using a single evaluation core, the calculations are performed by the main thread, otherwise it uses shared memory to transfer sampling points to the threads which can reduce the performance.

The following option is no longer supported: {\tt CubaCores}.

\subsection{Databases}

It is possible to work with databases containing integrals manually.
To do that one can use the {\tt kchashmgr} utility coming with the kyotocabinet database engine and the {\tt OutputIntegrand} tool coming as a part of {\tt FIESTA}.

To use {\tt kchashmgr} one first should run the 

{\tt . bin/lpath} 

command to add the proper path to {\tt LD\_LIBRARY\_PATH}.

Now it is possible to inspect the database with integrands. All calls contain {\tt -onl -onr} to prevent kyotocabinet from lengthy reorganization of the database.
First of all one should obtain the number of ``tasks''.

{\tt usr/bin/kchashmgr get -onl -onr get database-in.kch 0-}

Here and further {\tt database-in.kch} stands for the database name with full or relative path.
The result comes in a form like {\tt \{1, 2\}} listing integration tasks. For standard {\tt SDEvaluate} calls there is only one task numbered {\tt 1}, 
but for {\tt SDExpandAsy} there are more tasks due to region decomposition.

Now having a task number (we will use task {\tt 1} in further examples) one can obtain the number of sectors in this task by

{\tt usr/bin/kchashmgr get -onl -onr get database-in.kch \\ 1-SCounter}

and the list of ``prefixes'' --- the orders in $\ep$ and {\tt RegVar} that have to be calculated by 

{\tt usr/bin/kchashmgr get -onl -onr get database-in.kch \\ 1-ForEvaluation}

The number of sectors is a positive number and the list for example looks like 
{\tt \{\{-2, \{0, 0\}\}, \{-1, \{0, 0\}\}, \{0, \{0, 0\}\}, \{1, \{0, 0\}\}, \\ \{2, \{0, 0\}\}\}}.

Now it is possible to produce any integrand with a command looking like

{\tt bin/OutputIntegrand database-in.kch "1-2-\{0, 0\}-1000"}

Here {\tt 1} stands for the task number, {\tt 2-\{0, 0\}} for the prefix and {\tt 1000} for the sector number (additional ``{\tt-}'' symbols are separators).

A sector can contain multiple integrands that are distributed separately to integration workers with the {\tt SeparateTerms = True} option in {\tt Mathematica}
(or {\tt separate\_terms} for the pool binary). The number of such terms can be seen in the integral expression (as explained in the previous section) or obtained from the database with a command like

{\tt usr/bin/kchashmgr get -onl -onr get database-in.kch \\ "1-2-\{0, 0\}-1000"}

The result is an integer number. Now to obtain an integration term from {\tt 1} up to this number one can add the required number as the extra argument to the {\tt OutputIntegrand} call.

\section*{Acknowledgements}

I would like to thank M.~Steinhauser and V.~Smirnov for reading the draft of this paper as well as for constant support in my work. 
I an also thankful to T.~Hahn for his Cuba library and the readiness to improve it for the need of his users. 
I am also grateful to P.~Marquard for the numerous tests of the development versions of {\tt FIESTA}.

\section{Conclusion}

We presented a new release of {\tt FIESTA} --- a program for automatic numerical evaluation and analytic expansion of Feynman integrals.
This version has multiple improvements compared to the previous one resulting in an integration speed gain of about 2--4 times,
as well as a number of new useful options. Moreover this version can use graphical cards for calculations --- this can lead to
another 2--4 times speed improvement compared to the pure CPU usage.

\bibliographystyle{model1-num-names}
\bibliography{FIESTA4,asmirnov}
\end{document}